\documentclass[12pt]{article}
\newcommand{\be}{\begin{equation}}
\newcommand{\ee}{\end{equation}}
\newcommand{\bea}{\begin{eqnarray}}
\newcommand{\eea}{\end{eqnarray}}
\newcommand{\mr}{{\tilde{\mu}}}

\newcommand{\Ref}[1]{(\ref{#1})}
\newcommand{\ds}{/\hspace{-0.55em}\partial}

\newcommand{\pa}{\partial}

\newcommand{\la}{{\lambda}}

\newcommand{\oB}{|_{\partial M}}

\begin{document}

\noindent\today \hfill  YITP-02-13
\\
\vspace{2cm}
\begin{center} {\LARGE Heat kernels and zeta-function
regularization
for the mass of the supersymmetric kink}
\\  \vspace{2cm}
{Michael Bordag${}^*$\footnote{
Email: bordag@itp.uni-leipzig.de},  Alfred Scharff
Goldhaber${}^\dag$\footnote{Email:
goldhab@insti.physics.sunysb.edu},   Peter van
Nieuwenhuizen${}^\dag$\footnote{Email:
vannieu@insti.physics.sunysb.edu}, and
Dmitri Vassilevich${}^*$\footnote{Email:
Dmitri.Vassilevich@itp.uni-leipzig.de}\\[5pt]
${}^*${\it Institut f\"{u}r
Theoretische Physik, Universit\"{a}t Leipzig,
Augustusplatz 10,
D-04109, Leipzig, Germany}\\[5pt]
${}^\dag${\it C.N.Yang Institute for Theoretical
Physics, SUNY at
Stony Brook},
\\ {\it Stony Brook, NY 11794 } }
\abstract{ We apply zeta-function regularization to
the kink and
susy kink and compute its quantum mass.  We fix
ambiguities in the renormalization by considering
the asymptotic expansion of the vacuum energy as
the mass gap tends to
infinity (while keeping scattering data fixed)
and dropping all terms proportional to non-negative powers
or the logarithm of this mass gap. As an
alternative we write the regulated sum over zero point
energies in  terms of the heat kernel and apply
standard
heat kernel subtractions
with modifications which take the boundaries
into account. Finally we discuss to what
extent these procedures are equivalent to the usual
renormalization condition that tadpoles vanish.
}
\end{center}
\vfill\newpage

\section{Introduction}

In the 1970's and 1980's the problem of how to compute
quantum corrections to the mass of 1+1 dimensional
solitons received wide attention \cite{ref1}
\footnote{For general reviews of the quantization of
solitons
see \cite{Faddeev:1978rm}.}.
These soliton models could be made
supersymmetric (susy),  and
various methods to compute the quantum mass
were developed (see \cite{Schonfeld:1979hg}
and references in \cite{ref2}, \cite{Min} and
\cite{Graham:1999qq}).
It turned out that if one repeated for the susy kink
exactly
the same steps as performed in \cite{ref1} for
the bosonic case, the results for the quantum
mass of the susy kink depended both on the
regularization method and
the choice of boundary conditions.
In 1997, in view of renewed interest in
extended objects due to dualities in string theory,
the
issue of why different methods and different boundary
conditions for the quantum fields gave different
answers
for the quantum mass $M$
  (and the central charge $Z$ in the susy cases) was
reopened
\cite{ref2}.  It was found that two of the most
prominent
regularization schemes used  for these problems, mode
number cut-off and energy (=momentum)  cut-off,
gave different answers;  imposing different boundary
conditions yielded different answers as well.
As we shall review below, the reasons for these
differences
are by now well-understood, and both schemes yield the
same
correct answers provided they are applied correctly.

In this paper we apply another prominent
regularization scheme
to this problem: zeta function regularization
\cite{Dowker:1976tf}.
Our aim is not to settle the question what the correct
mass of
the kink  or susy kink is, but rather to show in a
clear and simple example
how to use zeta function regularization in problems
with boundaries.
We start from the regulated sum $\sum \omega_n^{1-2s}$
where
$s$ is a real positive number. To remove afterwards
the pole
$1/s$ we use two subtraction schemes often employed in
non-renormalizable field theories (note however that here
we are dealing with
a renormalizable theory).
The first scheme comes from 
Casimir energy calculations \cite{Bordag:2001qi}. One
just
subtracts the singularities (including the constant
term) in the
limit of large mass (to be discussed below).
  The second scheme,
widely used in quantum calculations on curved
backgrounds,
requires subtraction of the contributions from the
first
several heat kernel coefficients \cite{cbg}.
The heat kernel formalism has been used in quantum
theory since works by Fock \cite{11F} and Schwinger
\cite{11S}.
  Given
recent
advances of the technique
\cite{Gilkey,morehk} which take boundaries into account,
this scheme becomes a rather universal tool for
quantum
calculations in external fields. In this paper we
demonstrate
that the two schemes are equivalent: large mass
asymptotics is the same
as expansion in the proper time of the heat kernel.
Moreover,
in two-dimensional renormalizable field theories they
are
equivalent to the requirement that tadpoles are
absent, as
we prove in Appendix B.

Closer study of various regularization schemes for the
kink and the
susy kink in the period after 1997
revealed that the difference in answers was due
first of all to
  the fact that by evaluating sums over zero point
energies
one included  boundary energies for
most of the boundary conditions
considered \cite{Schonfeld:1979hg,Min}.  For particular boundary
conditions in the kink sector (the so-called twisted periodic and
twisted antiperiodic
boundary conditions) 
the boundary energy vanished, and
for these conditions mode regularization gave
the correct results \cite{Nastase:1999sy}.  The realization that for the
susy kink boundary energy should be subtracted suggested a
refinement of this scheme: one should average over
(quartets of) boundary conditions, so that upon
averaging the
  boundary energies would cancel
\cite{Goldhaber:2001ab}.
   For the energy
cut-off scheme, on the other hand, it was realized
that
instead of an abrupt cut-off at some maximal energy
one
should use a smooth cut-off, or the limit of a smooth cut-off,
and this yields an extra term which
resolved the problem
with the quantum mass both of the ordinary and of the
susy
kink \cite{LvNtbp}.
The mass of the susy kink has also been calculated
by scattering theory methods \cite{Graham:1999qq},
which
were applied to the
bosonic case earlier in \cite{Bordag:1995jz}.
(In
\cite{Bordag:1995jz}
even a renormalization prescription based on zeta
function regularization
was given. In this paper we shall justify this
prescription.)
Also some
other schemes for evaluating the sums over zero-point
energies were developed which remove boundary
energies,  in particular a method based on
first evaluating the $\partial / \partial m$
  derivative of the sum (where $m$ is the meson mass)
and
then integrating  this result w.r.t. $m$, using the
renormalization condition $M^{(1)}(m=0)=0$
\cite{Nastase:1999sy}.

For the reader who may wonder
whether there really is a problem in determining
the quantum correction to the kink mass by evaluation
of
$\sum\frac 1 2\hbar\omega_{n}$, we give an example which
should remove any doubt, and which also plays a role
below.  One can impose susy boundary conditions such
that
for every nonvanishing  bosonic solution
$\omega^{b}_{n}$
there is a corresponding fermionic solution with the
same
eigenvalue, $\omega^{f}_{n} = \omega^{b}_{n}$
(there are no bosonic or fermionic solutions with exactly zero energy
for the susy boundary conditions (\ref{set1})). The
one-loop correction to the susy kink mass is given by
\be\label{fsum} M^{(1)}=(\sum \frac 1 2 \hbar
\omega^{b}_{n}-\sum\frac 1 2\hbar\omega^{f}_{n})- (\sum \frac 1 2
\hbar
\omega^{b(0)}_{n}-\sum\frac 1 2\hbar\omega^{f(0)}_{n})+\Delta
M \ \ .
\ee Namely, one subtracts the contributions from the
trivial sector,  denoted by $(0)$, from those of the
topological sector,  and one must add a mass
counter\-term
$\Delta M$ , the finite part of which is fixed by
requiring the absence of tadpoles in flat space (or
far
from the kink). Clearly, all bosonic sums cancel
corresponding fermionic sums, but as $\Delta M$ is
nonvanishing and divergent,  even in the susy case,
this
procedure gives a divergent answer for $M^{(1)}$. One
can
explicitly evaluate the energy density near the
boundary
for susy boundary conditions, and finds then that it
is also
divergent \cite{Min}, giving as the boundary contribution $\Delta M-M^{(1)}$,
where
$M^{(1)}$ is the correct value of the mass of the
  susy kink \cite{GLvNtbp}. Subtracting the boundary
energy from the
total energy then gives the correct result.
  Thus the global answer actually is
correct for the entire system with susy boundary
conditions, but it is insufficient for obtaining the
mass of the
isolated kink.

By directly evaluating the local energy density one can
obtain the correct value for the quantum mass by
integrating this density around the kink. The first
such
method was developed in \cite{Min} where a
  higher-derivative action was used which preserves
susy
and is canonical  (in the sense of no higher time
derivatives). This scheme gave a result for the
quantum
mass which agreed with the results of Schonfeld
\cite{Schonfeld:1979hg},
the $\partial / \partial m$ method \cite{Nastase:1999sy}, 
and the phase shift methods \cite{Graham:1999qq}. 
For the susy sine-Gordon model, it also agrees
with the results obtained from the Yang-Baxter equation 
\cite{Nastase:1999sy}.
Another scheme to evaluate the local energy density
employs the concept
of local mode regularization \cite{Goldhaber:2001rp}.
This method also yields the correct result.

After this summary of previous work done on the
calculations of the
quantum mass of the susy kink,
we come to a new development: the application of
zeta-function regularization to this problem.  One
converts the sum
$\frac 1 2 \sum \hbar \omega_{n}^{1-2s}$ into an integral by
using a
function which has poles in $k$  corresponding to each
solution $\omega_{n}$. One can also convert each term
$\omega_{n}^{1-2s}$ in the sum into an integral by
using
the gamma function $\Gamma(s)$.  By writing the sum of
these integrals as  an integral over the sum, one
encounters the heat kernel, about which an enormous
literature exists.  We shall give a simple
self-contained
account directly applied to the case  of the kink and
susy kink.

In section 2 we discuss the bosonic kink with
Dirichlet boundary conditions and use the
large
mass subtraction scheme. In section 3 we consider the
susy
kink and use the heat kernel subtraction scheme. In
section 4 we summarize the methods. In Appendix A
we extend the large mass subtraction scheme of section 2
to Robin boundary conditions. As a by-product we derive
a simple relation between the contributions to the
vacuum energy from the boundaries and from the bound states.
In Appendix B we show that
these subtraction
schemes are equivalent to requiring absence of tadpoles.
\section{Large mass method for the bosonic kink}
The principle of zeta-function regularization of a sum
$\frac 1 2\sum \omega_{n}$ (we set $\hbar =1$) is to
replace this sum by
\begin{equation}\label{eq1}
E^{(0)}=\frac 12 \sum_{n}(\omega_{n}^{2} +
\mu^{2})^{1-2s} +
\frac 12 \sum_{i}(-\kappa_{i}^{2} + m^{2})^{\frac 1 2} \ \ ,
\end{equation}
  where $\omega_{n} = \sqrt{k_{n}^{2}+m^{2}}$  are the
zero point
energies for the continuum spectrum of the quantum fields
in a kink  background (discretized by putting the kink
system in a box with length $L$), and
$-\kappa_{i}^{2} + m^{2}=\omega^{2}_{B,i}$ are the
squared energies of the discrete spectrum, namely the
bound states and the zero mode. For the kink 
on an infinite interval there is
one bound state and one zero mode.  The real number
$s$ is  positive and large enough to
make the sum convergent, and
$\mu^{2}$ is a small mass temporarily introduced for
the treatment of the
zero mode (it will be used to move the pole
due to the zero mode
away from
the starting point of a branch cut).
Later we set $\mu$ to zero, and consider the
asymptotics of
$ E^{(0)}$ for $m\to \infty$ while keeping all
scattering data
$\kappa_i$ and
$k_n$ fixed (these scattering data may themselves depend on $m$). 
Our renormalization prescription will be
to
subtract all non-negative powers of $m$ in this limit
(together with $\ln m$
terms)\footnote{Inside the logarithm the mass $m$ must be divided
by some scale parameter $\tilde \mu$. Our subtraction procedure
does not depend on the choice of this scale parameter. Indeed,
if one chooses a different scale $\tilde{\tilde\mu}$ the logarithm
changes as $\ln (m/\tilde \mu)=\ln (m/\tilde{\tilde\mu})-
\ln (\tilde \mu/\tilde{\tilde\mu})$. The term on the l.h.s.\ of
this equation and the first term on the r.h.s.\  should be subtracted
since they are logarithmic in $m$. The second term on the r.h.s.
should be subtracted as well since it is proportional to $m^0$.
}. We shall see that this indeed gives a finite
result for
$ E^{(0)}$\footnote{The dependence of higher order correlation functions
(e.g., two-point functions) on the mass differs, of course, from
that of $ E^{(0)}$ due to the presence of extra dimensional
parameters (external momenta). The extension of the large mass subtraction
scheme to these correlation functions would require some extra effort. 
}. 

It is clear from (\ref{eq1}) that the zero mode ($\kappa^2=m^2$)
does not contribute to $E^{(0)}$.
For later use it is convenient to keep the zero mode
in (\ref{eq1}) since then some cancelations can be seen more
easily. However, we could drop the zero mode and still get the same
answer. Strictly speaking, there is not even a mode exactly at zero for
Dirichlet boundary conditions.
Sometimes, as in application of the heat
kernel technique (see sec. 3), both sums in (\ref{eq1}) must be 
considered on an equal footing. One then has to replace 
$(-\kappa_i^2 -m^2)^{\frac 12}$ by
$(-\kappa_i^2 -m^2)^{\frac 12 -s}$. This step is allowed in the
positive spectrum only. Therefore, in the next section
the zero modes will be excluded explicitly before applying the
zeta function regularization.

We first convert the continuum sum into an
integral
\begin{equation}
E^{(0)}_{\rm cont}=\frac 12 \oint\frac{dk}{2\pi
i}(k^{2}+M^{2}+\mu^2)^{\frac 1 2-s}
\frac {\partial}{\partial k}\ln \phi(k,m) \ \ .
\label{Econt}
\end{equation}
The function $\phi(k,m)$
has zeros at $k=k_{n}$ such that
$k^{2}_{n}+m^{2}=\omega_{n}^{2}$.
The
integration contour runs anti-clockwise and
consists of one branch at
$k=\Re k +i\epsilon$, a second branch at $k=\Re k
-i\epsilon$ ,
and a small segment
$-\epsilon \leq \Im k \leq \epsilon$ along the
imaginary
axis.

As the mass
parameter
$m$ may also appear in $\kappa_i$ and $k_n$, the
function $\phi(k,m)$
depends also on $m$. This $m$ should be kept fixed in
the asymptotic
limit. Therefore, we have replaced $m^2$ in
$\omega^2_n=k_n^2+m^2$ by
$M^2$. We also replace $m^2$ by $M^2$ in the sum over
bound states.
The asymptotic limit we need is $M\to\infty$.
The values of $\omega_n$ depend on the
boundary conditions, and the only place where boundary
conditions enter
in the calculations is in the choice of $\phi$.

We consider the supersymmetric Lagrangian
\begin{equation}
\label{sL}
{\cal
L}=-\frac12\left(\pa_\mu\Phi\right)\left(\pa^\mu\Phi\right)
-\frac12U^2(\Phi)-\frac12\overline{\psi} \left(\ds
+U'\right)\psi
\end{equation}
with polynomial superpotential
\begin{equation}
U=
\sqrt{\frac{\la}{2}}\left(\Phi^2-\frac{m^2}{2\la}\right)
\, \ \ .\label{U}
\end{equation}
The kink solution reads $\Phi =(m/\sqrt{2\lambda})
\tanh (mx/2)$. Here $m$ is the meson mass, and $\lambda$
is the (dimensionful) coupling constant.

In this section we consider the bosonic case, and impose
Dirichlet  boundary conditions\footnote{Robin boundary
conditions are considered in Appendix A.}
$\eta(-L/2)=\eta(L/2)=0$ on the scalar field
fluctuations $\eta$.
  Then
$\phi(k,m)$ is given by
\begin{equation}
\phi(k,m)=\sin (kL+\delta(k))
\label{fk}
\end{equation}
where the phase shift
$\delta(k)$ is given by \cite{ref1,Faddeev:1978rm}
\begin{equation}\label{deltak}
\delta(k)=-2 \arctan \frac{3mk}{m^{2}-2k^{2}}\,.
\end{equation}

Along the upper part of the contour we can approximate
$\sin (kL+\delta(k))$ by
$-\frac{1}{2i}\exp(-ikL-i\delta(k))$ because the term
with
$\exp(ikL+i\delta(k))$ vanishes as $L \rightarrow
\infty$.
Along the lower part of the contour we retain
$\frac{1}{2i}\exp(ikL+i\delta(k))$. We obtain then
\begin{equation}
E^{(0) I+II}_{cont}=\frac 12
\int\limits_{0}^{\infty}\frac{dk}{2\pi
i}(k^{2}+M^{2}+\mu^2)^{\frac 1 2-s}\frac{\partial}{\partial
k}[(ikL-\ln
e^{-i\delta(k)})+(ikL+\ln  e^{i\delta(k)})]\,.\end{equation}
The
contribution from the third part of the contour
is independent of the
phase shift, and hence it is dropped.
  Although $k=0$ is
formally a solution of the equation $\phi (k,m)=0$, the corresponding
mode is identically zero, and so the contribution from the pole at $k=0$
must be dropped; this is a peculiar property of Dirichlet boundary
conditions.\footnote{There is a careful way to exclude the pole
at $k=0$. 
In passing from the
sum over $n$ in Eq. \Ref{eq1} to the integral over $k$ in
Eq. \ref{Econt} one has the freedom to multiply $\phi(k,m)$ by a
function $f(k)$ without zeros and poles inside the contour whereby the
contour is initially chosen to intersect the real axis at
$k>0$. This function may be adjusted to give the product
$\phi(k,m)f(k)$ a finite nonzero value at $k=0$. As a result there
is no obstacle to moving the contour across $k=0$. The additional
contribution to the energy resulting from $f(k)$ is independent of the
background (in the case with $\delta(k)$ given by
Eq. (\ref{deltak}) one may take $f(k)=1/k$) and may be dropped.}
The terms
proportional to
$L$
are independent of the  phase shift and equal to the
contribution from the trivial vacuum, and as  usual
we
subtract them.
 
To
estimate the limit $M^{2} \rightarrow \infty$ of the
integral, it is useful to make a Wick rotation.
We replace $\delta(k)$ by
$-\delta(-k)$ in the second term because then the exponentials with
$\delta (k)$ become equal after the Wick rotation.
The
first part of the contour is rotated to the positive
imaginary axis  ($k=i\kappa$ with real positive
$\kappa$)
\begin{equation}
E^{(0) ,I}_{cont}-\mbox{L term}=\frac 12
\int\limits_{0}^{\infty}\frac{d\kappa}{2\pi
i}e^{i\pi(\frac 1 2-s)}[\kappa^{2}-M^{2}-\mu^2]^{\frac 1 2-s}
\frac{\partial}{\partial \kappa}[-\ln
e^{-i\delta(i\kappa)}]\,.\end{equation}
The second part of the
contour is
rotated downwards ($k=-i\kappa$, again with  real
positive
$\kappa$)
\begin{equation}
E^{(0) ,II}_{cont}-\mbox{L term}=\frac 12
\int\limits_{0}^{\infty}\frac{d\kappa}{2\pi
i}e^{-i\pi(\frac 1 2-s)}[\kappa^{2}-M^{2}-\mu^2]^{\frac 1 2-s}
\frac{\partial}{\partial \kappa}[\ln
e^{-i\delta(i\kappa)}]\,. \end{equation}
For $\kappa > M$ the sum of both terms contains the
factor
\begin{equation}
-e^{i\pi(\frac 1 2-s)} +e^{-i\pi(\frac 1 2-s)} = -2i\cos \pi s\,.
\end{equation}
For $\kappa < M$ the two integrals (almost) cancel
each other
so that only contributions of the poles in $\delta
(k)$ at imaginary $k$ need to be studied further.
\begin{eqnarray}&&\hspace{-10mm}
E^{(0) ,I+II}_{cont}-\mbox{L terms}=-\frac 12
\int\limits_{\sqrt{M^2+\mu^2}}^{\infty}\frac{d\kappa}{\pi}\cos\pi
s(\kappa^{2}-M^{2}-\mu^2)^{\frac 1 2-
s}[-i\frac{\partial}{\partial
\kappa}\delta(i\kappa)]\nonumber \\
&&\qquad\qquad\qquad\qquad\qquad
+{\mbox{pole contributions.}} \label{EE}
\end{eqnarray}
  We
now use an alternative form of the phase shift\footnote{
This is a particular case of the more general formula
(\ref{shiftgen}) valid for all reflectionless potentials.}
\begin{equation}
\delta(k)={i}\left[\ln \frac{k-im/2}{k+im/2}+\ln
\frac{k-im}{k+im}\right]\,.
\end{equation} It is clear from this
representation that the $\kappa$ integral has
poles at
$\kappa =\pm m/2$ and $\kappa =\pm m$. We now see the reason
for adding
$\mu^2$ to $m^2$ in (\ref{eq1}): it
  avoids that the starting point
of the cut at $k=iM$ coincides with the pole at $k=im$
(which is due
to the zero mode).
Each of the two contours
contains a half-circle  around these poles, but by
combining these two $\kappa$ contours from $0 \leq
\kappa \leq M$, one obtains the two residues at the
poles.
These two residues  precisely cancel the terms with
bound
states in the original sum.  
At this point we can put $\mu=0$.
Hence we  have arrived
at an
expression for the quantum mass as
an integral
over only
$M \leq \kappa \leq \infty$, and the bound state
contributions no longer
appear explicitly.\footnote{In other approaches one finds the
quantum mass expressed in terms of only the energies of the bound states;
see, e.g., \cite{Goldhaber:2001rp}.}

The expression for the quantum energy reduces to
\begin{equation}
E^{(0)}=-\frac 12
\int\limits_{M}^{\infty}\frac{d\kappa}{\pi}
(\cos\pi s)(\kappa^{2}-M^{2})^{\frac 1 2-s}
\frac{\partial}{\partial\kappa} \bigg{[}\ln
\frac{\kappa-m/2}{\kappa+m/2}+
\ln \frac{\kappa-m}{\kappa+m}\bigg{]} \,.\label{EEE}
\end{equation} 
We rewrite the expression in square
brackets as a sum of the limit of this  expression for
$\kappa \rightarrow \infty$ and a remainder
\begin{equation}
\ln \frac{\kappa-m/2}{\kappa+m/2}+
\ln \frac{\kappa-m}{\kappa+m}=
(-\frac{m}{\kappa}-\frac{2m}{\kappa})+ (\ln
\frac{\kappa-m/2}{\kappa+m/2}+
\ln \frac{\kappa-m}{\kappa+m}+\frac{3m}{\kappa})\,.
\end{equation}
With this decomposition inserted into the integral,
the first term is explicitly evaluated for
nonvanishing $s$,
whereas in the second term we may set $s=0$. 
The first, $s$-dependent, contribution is given by
\begin{eqnarray}
E^{(0)}_{(1)}&=&-\frac 12 \int_{M}^{\infty}
\frac{d\kappa}{\pi}
\cos\pi s(\kappa^{2}-M^2)^{\frac 1 2-s}
\frac{3m}{\kappa^{2}} \nonumber \\ 
&=&-\frac{\cos\pi
s}{4\pi}(3m)\, M^{-2s}
\left[ \frac{\Gamma(3/2-s)\Gamma(s)}{\Gamma(3/2)}
\right] \,,
\end{eqnarray}
while the second, $s$-independent, contribution yields
\begin{equation}
E^{(0)}_{(2)}=-\frac 12 \int\limits_{M}^{\infty}
\frac{d\kappa}{\pi}
\sqrt{\kappa^{2}-M^{2}} \left[ \frac{m}{\kappa^{2}-
m^{2}/4}+\frac{2m}{\kappa^{2}-m^{2}}-\frac{3m}{\kappa^{2}}
\right]\,.\label{eq17}
\end{equation}
For $s\to 0$ the term $E^{(0)}_{(1)}$ contains a pole
$1/s$ proportional to $M^0$
as well as finite contributions proportional to $M^0$
and $\ln M$.
According to our prescription this entire term must be
discarded.
The term $E^{(0)}_{(2)}$ vanishes in the limit
$M\to\infty$
and therefore must be kept in its entirety.  Direct evaluation of
this term
at $M=m$ yields the quantum mass of the bosonic
kink\footnote{It is convenient to represent the
numerator
of the 2nd term in square brackets in (\ref{eq17}) as $2m=-m+3m$
and then combine
these two parts with the 1st and the 3rd terms
respectively. Each of the
two integrals resulting from this procedure is
convergent and can be
easily calculated.}
\begin{equation}
M_b=m \left( \frac 1{4\sqrt{3}} -\frac 3{2\pi} \right)\,.
\label{Mb}
\end{equation}
This is the correct result
\cite{ref2,Min,Graham:1999qq}
first found in \cite{ref1}.  

Although the terms in the square brackets in (\ref{EEE})
are related to two bound states on the background of the bosonic kink,
these two terms do not describe contributions of from the modes of the
discrete spectrum to the vacuum energy. The expression (\ref{EEE})
has been obtained by a contour rotation. During this rotation, the
pole terms (which are genuine contributions of the discrete spectrum)
were cancelled by a part of the continuous spectrum. 
Note that both terms in (\ref{EEE}) are divergent while contributions
of the bound states to $E^{(0)}$ are always finite and do not even
require a regularization (see eq. (\ref{eq1}).

\section{Heat kernel method for the susy kink}
We now consider the susy kink.  The classical kink
solution satisfies $\partial_1\phi_K+U =0$, and
the background solution $\phi =\phi_K$ and $\psi =0$
is invariant under rigid susy
with parameter $\epsilon_-$, as\footnote{We set
$\psi =\begin{pmatrix}{
\psi_+ \cr \psi_-}\end{pmatrix}$ and 
$\epsilon =\begin{pmatrix}{\epsilon_+ \cr \epsilon_-}\end{pmatrix}$
and use $\gamma^1=
\begin{pmatrix}{1&0\cr 0& -1}\end{pmatrix}$
and $\gamma^0=
\begin{pmatrix}{0& -1\cr 1 &0 }\end{pmatrix}$.}
\begin{eqnarray}
&&\delta\psi_+=\partial_1\phi\epsilon_+-\partial_t\phi
\epsilon_--U\epsilon_+ \,,\nonumber \\
&&\delta\psi_-=-\partial_1\phi\epsilon_-+\partial_t\phi
\epsilon_+-U\epsilon_- \,.
\end{eqnarray}  Note also 
$\delta\phi =-i\epsilon_+\psi_-+i\epsilon_-\psi_+$. There are
two sets of supersymmetric boundary conditions
for the quantum fields which are invariant under the $\epsilon_-$
transformations.
The first
set reads
\begin{eqnarray}
&&\eta\oB =0,\qquad \psi_+\oB =0 \,,\nonumber \\
&&(\partial_1 -U'(\phi ))\psi_- \oB =0 \ \ .
\label{set1}
\end{eqnarray}
The second set is
\begin{eqnarray}
&&(\partial_1 +U'(\phi ))\eta \oB =0,\qquad \psi_-\oB
=0\,, \nonumber \\
&&(\partial_1 +U'(\phi ))\psi_+ \oB =0 \ \  .
\label{set2}
\end{eqnarray}
For the spinor field it is sufficient
to impose the Dirichlet boundary conditions for one
component.
The Robin boundary conditions for
the other component are then determined by the Dirac
equation, which is equivalent to the following equations
on the spinor field components
\begin{eqnarray}
&&\left( \partial_1+U'\right)\psi_+ -\partial_0\psi_-=0 \, \ \ ,
\label{D1} \\
&&\left( -\partial_1+U'\right)\psi_- +\partial_0\psi_+=0 \, \ \ .
\label{D2}
\end{eqnarray}
If we require that the $\psi_+$ component satisfy the Dirichlet
boundary condition (the 2nd equation in (\ref{set1})), 
equation (\ref{D2}) then yields the Robin boundary condition
for $\psi_-$ (2nd line in (\ref{set1})). The same is also true
for (\ref{set2}).
We need these boundary conditions for the other spinor
component as well because the heat
kernel method uses second-order differential
operators, so that
we must exclude spurious solutions\footnote{The susy invariance of 
(\ref{set1}) under $\epsilon_-$ transformations is easy to prove;
in particular the last condition in (\ref{set1}) transforms into
the full $\phi$ field equation which becomes $\partial /\partial t$ independent
at the boundary due to the condition $\eta =0$, and therefore this
field equation itself is a boundary condition \cite{RPvN}. The first
of the second set of boundary conditions should read 
$\partial_1\phi +U=0$. Then the first condition transforms into the 
second, the second into the third, and the third into
$(\partial_1 +U')\dot\eta =0$. The last condition follows 
from the first by taking the time derivative.}.

To discuss the cancelation of the bosonic and fermionic
contributions
to the vacuum energy, it is convenient to introduce
the operators
\be\label{Q+-}Q_\pm=\pa_x\pm U'
\ee
and build from them the second-order differential
operators
\begin{equation}
\Delta_-=Q_+Q_-\,,\quad
\Delta_+=Q_-Q_+\, \ \ . \label{Qpm}
\end{equation}
The eigenfrequencies of the bosonic fluctuations
and those of $\psi_+$ are defined by the operator
$\Delta_+$, but the
eigenfrequencies for $\psi_-$ are defined by
$\Delta_-$.
The operators $\Delta_+$ and $\Delta_-$ satisfy the
following
intertwining relations
\begin{equation}
\Delta_-Q_+=Q_+\Delta_+\,,\qquad
Q_-\Delta_-=\Delta_+Q_-\, \ \ .
\label{inter}
\end{equation}
Moreover, the boundary conditions (\ref{set1}) and
(\ref{set2})
are compatible with these relations: if $\psi_+$ is an
eigenfunction
of $\Delta_+$ which vanishes at the boundary, then
$Q_+\psi_+$ is an
eigenfunction of $\Delta_-$ which obeys the second
condition in (\ref{set1}).
Conversely, if $\psi_-$ is an eigenfunction of $\Delta_-$
which satisfies the second condition in (\ref{set1}), then
$Q_-\psi_-$ is an eigenfunction of $\Delta_+$ which vanishes
at the boundary.
Therefore,
the two operators involved are isospectral (up to the
zero modes which
do not contribute to the vacuum energy). This shows
a complete compensation of the regulated sum of the
zero
point energies
in the bosonic
and fermionic sector.

The property of isospectrality is crucial in our
discussion.
Nonsupersymmetric boundary conditions in principle
also can be
handled with heat kernel methods, but the treatment is
more complicated.

Each operator $\Delta_\pm$ can be represented as
$\Delta_\pm =\partial_1^2 -m^2 -V_\pm$, where $m^2$ is
chosen in such a
way that $V_\pm$ vanishes as $x^1\to\pm\infty$. The
potentials then read
$V_{\pm}=U'^2\pm U''U-m^2$. From now on we drop the
subscript $\pm$
where this cannot lead to confusion.
Let $k_n^2+m^2$ be the eigenvalues of $\Delta$.
In zeta functional regularization the regularized
vacuum energy
has the form
\be\label{E0}E_0=\frac{\mr^{2s}}{2}\sum_n
\left(k_n^2+M^2\right)^{\frac12-s} \ \ ,  \label{sumen}
\ee
where $\mr$ is an arbitrary parameter with the
dimension of a
mass which is needed to restore the correct dimension
of $E_0$.
Again, we have replaced the mass $m$ in the spectrum by $M$
to
be able to consider later the limit $M\to \infty$
without deforming
the potential $V$.
The mass parameter
  $m$ which enters the susy boundary conditions (\ref{set1})
  and (\ref{set2}) through the superpotential $U$ also must
  be kept fixed in the limit $M\to\infty$. The values of $k_n$
  do not depend on $M$ and are the same in the fermionic and
bosonic sectors (up to the zero mode which must be excluded
  explicitly\footnote{
If one writes the zero mode contribution as square root of $-m^2+M^2$,
taking the limit $M\to\infty$, the zero mode would contribute
erroneously.}). Therefore, the bosonic
  contribution to the vacuum energy cancels the fermionic
  contribution for all values of $M$. Consequently, in the
large $M$ subtraction scheme both $E_0$ and $E_0^{\rm div}$ vanish for
the susy kink. This gives the value zero for the  total energy of the susy
kink and the boundaries. As we  shall see below, this is the correct
result.
  
To introduce the heat kernel subtraction scheme
  we first consider an individual contribution (bosonic or
  fermionic) to the regularized vacuum energy.
  We write
\be\label{e->k}
\sum_n \left(k_n^2+M^2\right)^{\frac12-s}
=
\int_0^\infty\frac{dt}{t}\frac{t^{-\frac12+s}}{\Gamma
\left(s-\frac12\right)} \ K(t) \ e^{-tM^2} \ \ \ ,
\ee
where
\be\label{hk}K(t)=\sum_n \ e^{-tk_n^2}
\ee
is the heat kernel for this problem. Its asymptotic
expansion for $t\to0$,
\be\label{hke}K(t)\sim \sum_{n\ge 0} a_n t^{n-\frac
12} \
\ee
is equivalent to the expansion of $E_0$ for
$M\to\infty$. The ultraviolet
divergences are contained in the contributions with
$n=0,\frac12,1$
which correspond to the nonnegative powers of $M$.
{}From this
we define the divergent part of the vacuum energy as
$$E_0^{\rm
div} \equiv \frac{\mr^{2s}}{2}\int_0^\infty
\frac{dt}{t}\frac{t^{-\frac12+s}}
{\Gamma\left(s-\frac12\right)}
\sum_{n= 0}^1 a_n t^n \ e^{-tM^2} = $$
\bea\label{Ediv}\frac{\mr^{2s}}{2\Gamma\left( s-\frac 12 \right)}
\left\{
a_0 \Gamma \left( s-1\right ) M^{2-2s} +
a_{\frac 12} \Gamma \left( s-\frac 12 \right )
M^{1-2s} +
a_1 \Gamma \left( s\right ) M^{-2s} \right\}  .
\eea
We define the renormalized vacuum energy as
\be\label{Eren}E_0^{\rm ren}=E_0-E_0^{\rm div}
\ee
at $s=0$ and  $M=m$. For $n$ larger than unity,
the coefficients $a_n$ are multiplied by negative
powers of
$M$, so that the subtraction
of the leading heat kernel asymptotics for small $t$
is equivalent
to the subtraction of the leading large mass
asymptotics. In
Appendix B it is demonstrated that for renormalizable
theories these two
procedures are equivalent to imposing the condition
that
tadpoles vanish.

To obtain the kink mass, we must determine the
contributions to the
heat kernel from the boundary. This requires a
detailed
discussion of the heat
kernel coefficients\footnote{
The expansion (\ref{hke}) has quite recently been used
in numerical
calculations of the kink mass
\cite{Izquierdo:2002eb}.}.
To write down these coefficients we need
some new notations.
Let us introduce the boundary operators ${\cal
B}^{D,R}$ acting on a
field $\varphi$ (either bosonic or fermionic)
\begin{equation}
{\cal B}^D\varphi = \varphi \, \ \ ,\qquad {\cal
B}^R\varphi =
(\partial_N+S)\varphi \, \ \ .\label{bop}
\end{equation}
Here $\partial_N$ denotes partial derivative with
respect to an inward
pointing unit vector. The field $\varphi$ satisfies
the field equation
$(-\partial_x^2+V)\varphi_n =\epsilon_n^2\varphi_n$.

The asymptotic expansion of the heat kernel $K(t)$ as
$t\to 0$ for the boundary
condition ${\cal B}\varphi \oB =0$ (where
${\cal B}$ can be ${\cal B}^D$ or ${\cal B}^R$) reads
\begin{equation}
K(t)=\sum\limits_{n=0,\frac 1 2,1,\dots} t^{n-\frac 1 2}a_n({\cal
B}^{R,D} ) \ \ ,
\label{asex}
\end{equation}
with \cite{Gilkey}
\begin{eqnarray}
a_0({\cal B}^{R,D} )&=&(4\pi )^{-\frac 1 2}\int_M
dx\, \ \ ,\nonumber \\
a_{\frac 1 2} ({\cal B}^{R} )&=& \frac 14 \int_{\partial M}
dx \, \ \ ,
\qquad a_{\frac 1 2} ({\cal B}^{D} )=- \frac 14
\int_{\partial M}  dx
\, \ \ ,\nonumber\\
a_1({\cal B}^{R} ) &=&(4\pi )^{-\frac 1 2}\left( -\int_M V
dx +
2\int_{\partial M} S dx \right) \ \ , \nonumber \\
a_1({\cal B}^{D} ) &=&-(4\pi )^{-\frac 1 2} \int_M V dx\, \ \ .
  \label{a}
\end{eqnarray}
In the present (one-dimensional)
case the boundary integral becomes a sum over the two
boundary points.

For the fermionic sector we sum the contributions from
$\psi_+$
and $\psi_-$, and divide by a factor $2$.
It is easy to see that in the ``total'' heat kernel
\begin{equation}
a_n^{tot}=a_n^{bos}-a_n^{ferm} \label{atot}
\end{equation}
  the contribution of $\psi_+$ always
cancels one half the contribution of $\eta$.

Let the kink be centered at the origin, but the box
lie
between $x=L_1$ and $x=L_2$.
In all supersymmetric theories we have
$a_0^{tot}=0$ independently of the background and the
boundary
conditions. This is due to the equality of the
numbers of bosonic and fermionic
degrees of freedom.
The coefficient $a_{\frac 1 2}$ simply counts the number of
Dirichlet minus Robin boundary conditions in each set
(\ref{set1}) or
(\ref{set2}). Obviously,
\begin{equation}
a_{\frac 1 2}^{tot}=-\frac 12 \label{a121}
\end{equation}
for the boundary conditions (\ref{set1}), and
\begin{equation}
a_{\frac 1 2}^{tot}=\frac 12 \label{a122}
\end{equation}
for the boundary conditions (\ref{set2}). However, we
must set
$a_{\frac 1 2}^{tot}=0$ for the following reason\footnote{
This follows also from the isospectrality arguments (see above).
}. The operator
$\Delta_+$ (resp.
$\Delta_-$) has a zero mode which satisfies
the Robin boundary condition
$Q_+\psi_+=0$ (resp.
$Q_-\psi_-=0$).
Therefore, there is one zero mode in the fermionic
sector
for the boundary conditions (\ref{set1}).  (There are of
course no zero
modes in the bosonic sector because we use
Dirichlet boundary
conditions there\footnote{The zero mode $\eta_0$ in the bosonic
sector is a solution of the equation $Q_+\eta_0=0$. Explicitly,
$\eta_0=c\partial_1\phi_K$, where $c$ is a constant and $\phi_K$
is the kink solution. On an infinite space the function $\eta_0$
is indeed a normalizable zero mode. However, for finite $L$
the Dirichlet
boundary condition $\eta_0\vert_{x=\pm L/2}$ yields $c=0$.
A more detailed analysis shows that on a finite interval
this mode is shifted and receives a positive energy.
Note, that there was no zero mode
contribution in the calculations of the previous section.
}).
Each bosonic (or fermionic) zero mode contributes $+\frac 12$ (or
$-\frac 12$) to the heat kernel coefficient. 
The zero mode structure, being a global characteristic of the problem,
depends crucially on the boundary conditions 
(cf. \cite{Graham:1999qq,Graham:1999pp,Goldhaber:2001ab} where the zero
mode structure for other boundary conditions has been analysed).
When the contributions from the boundaries to the vacuum energy are
properly subtracted the mass of an isolated kink must not depend on
the boundary conditions. In the zeta function regularization,
zero mode
contributions must be
explicitly subtracted from the vacuum energy. A
similar mechanism works
also for the boundary conditions (\ref{set2}).

Let us now turn to calculation of $a_1$.
The volume terms in $a_1$ combine to give
\begin{eqnarray}
a_1^{tot}[\mbox{vol}]&&=\frac 12 (4\pi )^{-\frac 1 2}\int_M
(-V_++V_-) dx\nonumber \\
&&=\frac 12 (4\pi
)^{-\frac 1 2}\int\limits_{L_1}^{L_2}(-2UU'') dx\nonumber \\
&&=\frac 12 (4\pi )^{-\frac 1 2} (-2U'(L_1)+2U'(L_2))
\label{a1vol} \ \ ,
\end{eqnarray} 
where we have used the Bogomolny equation $\partial_1
\phi =-U(\phi )$.

We next calculate contributions to $a_1$ from
$\partial M$.
We have to
cast the boundary conditions into the standard form
(\ref{bop}).
Note that $\partial_1=\partial_N$ at the left
boundary, but
$\partial_1=-\partial_N$ at the right boundary.
Consequently, we have to
set $S=-U'$ at $x=L_1$ and $S=U'$ at $x=L_2$ for
$\psi_-$ and the
boundary conditions (\ref{set1}). For the boundary
conditions
(\ref{set2}) $S$ assumes the opposite
values, but now both $\eta$ and $\psi_+$ contribute.
As a result, we obtain for (\ref{set1}) and
(\ref{set2}) equal
boundary contributions
\begin{equation}
a_1^{tot}[\mbox{boundary}]=\frac 12 (4\pi )^{-\frac 1 2}
(2U'(L_1)-2U'(L_2))
\label{a1bou}
\end{equation}
which exactly cancel the volume term (\ref{a1vol})
giving
\begin{equation}
a_1^{tot}=0 \,\label{a1tot} \ \ .
\end{equation}
We conclude that all singularities as well as the
finite part of the
contributions to the vacuum energy cancel. Thus,
according to (\ref{Eren}),
not only $E_0$ but also
the renormalized energy $E_0^{\rm ren}$ of the kink
plus boundaries vanishes.

Qualitatively, it is clear what has happened. The
vacuum energy
associated with the kink
has been cancelled by the vacuum energy associated
with the boundaries.
Therefore, to obtain the correction to the mass of an
isolated kink we have
to subtract from the above vanishing result the
Casimir energy of the
boundaries as they are being moved to infinity.

Of course, the kink solution cannot be smoothly
deformed to the trivial
one. We can, however, replace all background fields
in the wave operator for the scalar field
fluctuations and in the squared Dirac operator by
their asymptotic values
($V=0$), and replace $U'$ in (\ref{set2}) by $+m$ at
$x=L/2$ and
$-m$ at $x=-L/2$.
In this way, we arrive at the problem of calculating
the vacuum
energy for free scalar and spinor fields with mass
$m$.
We consider the boundary
conditions (\ref{set2}) because, as we shall later
explain,
the calculations with (\ref{set1}) are more
complicated.
The boundary conditions for the free fields read
\begin{eqnarray}
&&(\partial_N -m)\eta \oB =0,\qquad \psi_-\oB =0\,,
\nonumber \\
&&(\partial_N -m)\psi_+ \oB =0 \,, \label{set2f}
\end{eqnarray}
where $\partial_N=\partial_1$ on the left boundary
$x^1=-L/2$, and
$\partial_N=-\partial_1$ at $x^1=L/2$. Obviously, the
bosonic and fermionic
contributions partially cancel each other. For
(\ref{set2f}) the vacuum
energy $E^{[B]}$ associated with the boundaries become
\begin{equation}
E^{[B]}=\frac 12 (E^{[R]}-E^{[D]}),
\label{EB}
\end{equation}
where the superscripts $[R]$ and $[D]$ stand for Robin
and Dirichlet
boundary conditions respectively.

Let us start with the Robin sector.
The eigenfunctions of the operator $-\partial_1^2
+m^2$
\begin{equation}
\varphi_k=A_k \sin (kx) +B_k \cos (kx)
\label{eigf}
\end{equation}
satisfy the Robin boundary conditions in (\ref{set2f})
if
\begin{equation}
k \cos \left( \frac{kL}2 \right) + m \sin \left( \frac{kL}2 \right)
=0\quad \mbox{and}\quad
B_k=0 \label{1stsol}
\end{equation}
or if
\begin{equation}
-k \sin \left( \frac{kL}2 \right) + m \cos \left( \frac{kL}2 \right)
=0\quad \mbox{and}\quad
A_k=0 \label{2ndsol} \ \ .
\end{equation}
It is easy to
see that the two equations (\ref{1stsol}) and
(\ref{2ndsol}) on the wave numbers $k$ in
are equivalent to a single equation,
\begin{equation}
f_R(k)=\sin \left( 2\left(  \frac{kL}2 
  +\delta_R(k)\right) \right) =0 \quad
{\mbox{with}}\quad \
\delta_R (k)=\arctan (k/m)
\label{spectrum} \ \ .
\end{equation}
We can use the integral representation (\ref{Econt})
for the regularized vacuum energy
\begin{equation}
E_0^{[R]}=\frac {\mr^{2s}}{2}
\oint \frac {dk}{2\pi i} \left( k^2 +m^2
\right)^{\frac 12 -s} \frac{\partial}{\partial k}\ln
f_R(k)  \ \ ,
\label{ERob}
\end{equation}
where the contour again encircles the real positive
semi-axis.

Clearly, the same representation is also valid for
$E_0^{[D]}$
with $\delta_D(k)=0$.
The contour rotation proceeds in exactly the same way
as before
(although there are no bound states in this case).

After rather elementary (but
somewhat lengthy) calculations
we obtain in the limit $L\to\infty$
\begin{equation}
E^{[B]}(s)=\frac 12 \left( \frac {\mr^{2s} \Gamma (s)
m^{1-2s}}{2\sqrt{\pi}
\Gamma \left( s+\frac 12 \right)} +\frac 12 m \right)
+\dots
\label{EBs}
\end{equation}

Again we make the heat kernel expansion of the energy
in the trivial
sector (which is only located on the boundaries).
Because the boundary conditions (which were susy in
the kink sector)
break susy in the trivial sector, both $E_0$ and
$E_0^{\rm div}$ are
nonvanishing in this case. From the equations
(\ref{a})
we immediately obtain
\begin{equation}
a_0^{[B]}=0,\quad a_{\frac 1 2}^{[B]}=\frac 12 ,\quad
a_1^{[B]}=-\frac m{\sqrt{\pi}}
\,.\label{aB}
\end{equation}
These coefficients define the divergent part of the
vacuum energy (\ref{Ediv}).
The boundary energy is the difference of the total
result in the boundary
energy in (\ref{EBs}) and the divergence in
(\ref{Ediv})
\begin{equation}
E^{[B]{\rm ren}}=\frac m{2\pi} \,. \label{EBren}
\end{equation}
The one-loop correction $M^{(1)}$to the kink mass is
the difference between
the one-loop vacuum energy of the kink on a finite
interval (which is
zero) and the vacuum energy (\ref{EBren}) associated
with the
boundaries. Thus we obtain
\begin{equation}
M^{(1)}=-\frac m{2\pi} \,. \label{M1}
\end{equation}
This is the correct result.

Direct application of the same method to the other set
of the boundary
conditions (\ref{set1}) yields in addition to the
eigenmodes (\ref{eigf})
for $\psi_-$ also the eigenmodes of the form $A_k
\sinh (2kx/L)+
B_k \cosh (2kx/L)$ which correspond to negative
eigenvalues of
$-\partial^2_1$. Their absolute values are large
enough to make
$-\partial_1^2 +m^2$ negative. In this
case one cannot replace $U'(\phi )$ in the boundary
conditions by
its asymptotic values $\pm m$ before taking the limit
$L\to\infty$.
Keeping $L$-dependent boundary conditions considerably
complicates
the calculation. This case is considered from a little
bit different point of view in Appendix A.

\section{Conclusions}
We have illustrated in a simple concrete physical
model how one may
use zeta function regularization and heat kernel
methods. The problem was
not totally trivial because there were boundaries
present which played
a role in the trivial and the topological sector.

In the zeta function approach, one converts the sum of
the modes into a
complex contour integral. The renormalization
condition which fixes the
finite part of the zero point energy is that one must
discard all terms proportional to non-negative powers
of $M$ or $\log M$ 
in the limit when the meson mass $M$ tends to
infinity while keeping the ``scattering data'' (namely
$k_i$ and $\kappa_i$)
fixed. In our case these terms were proportional to
$M^0$ and $\ln M$.
The remainder gave the correct mass of the
bosonic kink.

In the approach which uses heat kernels, the sum over
modes was converted
into an integral over the proper time $t$, and the
asymptotic expansion
in $t$ yielded the heat kernel coefficients $a_n$. We
first computed the total
energy in the susy kink sector, and defined its
renormalization as
subtraction of the contributions from $a_0$, $a_{\frac 1 2}$
and $a_1$. Both the total and renormalized energy were
found to vanish in the topological sector.
This was to be expected because we used susy boundary
conditions in the kink sector, so that the sum over
bosonic modes should exactly cancel the sum over fermionic
modes.
  Then we repeated
this calculation
in the trivial sector, but with the same boundary
conditions as for
large $x$ in the topological sector.  These boundary
conditions are not susy in the trivial sector.  As a
result we found that the total energy and the
renormalized energy were nonvanishing. 
The difference of the renormalized energy in the topological sector
and in the trivial sector (which is thus minus the renormalized
energy in the trivial sector in this case) is the kink mass.
This difference
yielded the correct result for the susy kink. A
nontrivial technical aspect of this approach
was the role of Robin boundary conditions and zero
modes in the heat
kernel expansion.

The physical meaning of subtracting the contribution of the heat
kernel coefficient $a_0$ is to subtract the energy of the trivial
vacuum. However, the physical meaning of subtracting the contribution
of the heat kernel coefficient $a_1$ is twofold: as expected it
subtracts the tadpole contribution (the contribution from the counter
term) and it adds the anomaly. It is well known that in
scale invariant theories $a_1$ contains the trace anomaly; 
here it contains the
anomaly found in \cite{Nastase:1999sy,Min}. We shall 
discuss this elsewhere in more detail.

Of course the
problem confronting any global method using fixed boundary conditions in
the kink and the trivial sectors, that there is a boundary energy for
fermions in addition to the kink energy, also confronts heat kernel 
regularization.
The method we use to overcome this problem, namely creating a new
problem in which only the boundary energy is present, presumably could
be used in any regularization scheme.  In particular, we have checked
that it works for zeta function regularization with large-mass subtraction.
We stress that the large mass subtraction scheme is fully equivalent
to the heat kernel subtraction. It is possible, of course, to use
the heat kernel subtraction in the bosonic case and the large mass
subtraction for the fermions.
\section*{Acknowledgements}
This work has been
supported in part by the
Deutsche Forschungsgemeinschaft (project BO
1112/11-1), Erwin Schr\"{o}dinger
Institute (Vienna) , the Austrian Science Foundation
(project P-14650-TPH), and the NSF (project PHY-0071018).
\section*{Appendix A: Large mass subtraction scheme for
Robin boundary conditions}
In this Appendix we demonstrate how the large $M$ subtraction
scheme works for generic Robin boundary conditions
(Robin boundary conditions are a linear combination of Dirichlet
and Neumann boundary conditions).
All necessary
information on one-dimensional scattering theory can be
found in \cite{Chadan}. Consider a reflectionless
potential $V(x)$ localized around the point $x=0$
(and vanishing for $x\to\pm\infty$).
Before imposing any boundary condition we have two independent
eigenfunctions $\varphi_{1,2}(k,x)$ of the operator
$\Delta =\partial_1^2-m^2-V(x)$
for each $k$. Their asymptotics
as $x\to \pm\infty$ are
\begin{equation}
\varphi_1(k,x)\to \cos (kx \pm \delta/2) \ \ , \qquad
\varphi_2(k,x)\to \sin (kx \pm \delta/2) \ \ , \label{asympt}
\end{equation}
where $\delta(k)$ is the phase shift defined by the potential $V$.
We put the system in the box $-L/2 \le x \le L/2$ and suppose
that $L$ is large enough
  so that near the boundaries
the eigenfunctions can be represented by their asymptotics.

Consider the Robin boundary conditions
\begin{equation}
\left. (\partial_x +S_1)\varphi\right|_{(x=-L/2)}
= \left. (-\partial_x +S_2)\varphi\right|_{(x=L/2)}=0
\label{gRbc}
\end{equation}
with arbitrary $S_1$ and $S_2$. The
general eigenmode of the operator
$\Delta$ is a linear combination of $\varphi_1$ and $\varphi_2$:
$\varphi (k,x)=A\varphi_1(k,x)+B\varphi_2(k,x)$. Substituting
the asymptotics (\ref{asympt}) of this eigenmode
in the boundary condition (\ref{gRbc})
we arrive at the following equation which defines the spectrum of $k$:
\begin{equation}
0=\phi (k)=\sin (kL+\delta +\alpha_1 +\alpha_2) \,
\ \ ,\label{newshift}
\end{equation}
where
\begin{equation}
\alpha_{1,2}=-\arctan (k/S_{1,2}) \, \ \ .\label{alpha}
\end{equation}

The vacuum energy is now defined by the equation (\ref{Econt})
where we have to substitute the function $\phi (k)$ given
in (\ref{newshift}) instead of $\phi (k,m)$. If we now repeat the
same steps as in sec. 2 we arrive at the equation (\ref{EE})
where $\delta$ must be replaced by $\delta +\alpha$,
$\alpha =\alpha_1+\alpha_2$. The
important property of the equation (\ref{EE}) is the linearity
in $\delta$. Therefore, the regularized vacuum energy for
arbitrary auxiliary mass $M$ can be represented as a sum of
the contribution from the potential $V$ (containing $\delta$
under the integral) and the boundary contribution (containing
$\alpha$). To obtain the mass of the kink alone we have to subtract the
whole boundary contribution, retaining the $\delta$-term only.
Hence we arrive at the old expression for the kink mass obtained
for the Dirichlet boundary conditions. We conclude, that if the
boundary contributions are subtracted the large $M$ scheme gives
equivalent results for the kink mass independently of the
boundary conditions.

We conclude this Appendix with a useful relation between the
contributions to the vacuum energy from the bound states
and from the boundaries. For any
reflectionless potential the phase shift can be represented
as a sum over the bound states \cite{Chadan}
\begin{equation}
\delta (k) =i\sum\limits_n \ln \frac {k-i\kappa_n}{k+i\kappa_n}=
\sum\limits_n 2\arctan \left( \frac {\kappa_n}k \right) \,,
\label{shiftgen}
\end{equation}
where $\kappa_n$ are the bound state momenta. We can rewrite
(\ref{alpha}) in a similar form
\begin{equation}
\alpha =\alpha_1 +\alpha_2=
\arctan \left( \frac {S_1}k \right) +
\arctan \left( \frac {S_2}k \right) -\pi \, \ \ .\label{newa}
\end{equation}
Because of the derivative $\partial_k$ in (\ref{EE}) the constant term
$-\pi$ in (\ref{newa}) does not contribute to the vacuum energy. By 
comparing
(\ref{shiftgen}) and (\ref{newa}) we see that the vacuum energy
associated with
a boundary with Robin boundary conditions\footnote{
In the large $M$ subtraction scheme the vacuum energy
associated with the Dirichlet boundaries is zero. This
is  clear from the considerations of sec.~2. The reason is
that the Dirichlet problem contains no dimensional parameter.
Therefore, the boundary energy can only be proportional to
$M$. In the large $M$ scheme such contributions must
be dropped.}
  is,
roughly speaking, one half
of the contribution of a bound state with the energy $\kappa =S$.

As an example, consider the supersymmetric boundary conditions
(\ref{set1}). The bosonic sector is described by the potential
$V_+$ and contains two bound states with the energies
$\kappa_1=m$ and $\kappa_2=m/2$. There is no boundary
contribution to the vacuum energy. The same is true for $\psi_+$.
The other spinor component $\psi_-$ corresponds to the potential
$V_-$ and has one bound state with $\kappa_2=m/2$.
For large $L$ the Robin boundary conditions have $S_1=S_2=m$.
These two boundaries add the same
contribution to the vacuum energy as would come from the bound state
with $\kappa_1=m$. This restores the balance between the bosonic and
fermionic  contributions and gives zero total energy for the susy kink
and the boundaries. A similar mechanism works for the
boundary conditions (\ref{set2}), where $S_1=S_2=-m$ and
the boundary contribution is minus the contribution
of the bound state $\kappa_1$.
\section*{Appendix B: Equivalence of the large mass and heat
kernel subtraction schemes to the vanishing tadpole condition}

In this Appendix we demonstrate that in a
renormalizable theory
in two dimensions the heat kernel subtraction
procedure
is equivalent
to imposing the renormalization condition of vanishing
tadpoles in the trivial
vacuum. As the former scheme is equivalent to the large
$M$ subtraction (see sec.~3), this will also prove that all
three schemes are equivalent.
For simplicity, we restrict ourselves to a
bosonic theory.

Consider first the case of a constant but arbitrary
background field $\Phi$.
The second-order differential operator acting on the
fluctuations
has the form $-\partial_1^2 +M(\Phi )^2$. For the
action (\ref{sL})
$M^2(\Phi )=U'(\Phi )^2 +U(\Phi )U''(\Phi )$. The
explicit form of $M^2$
will play no role. The complete heat kernel in this
case  is given
by the term with $n=0$ in
(\ref{hke}). As this term is proportional
to the volume of the manifold, it is more convenient
to consider the
energy density $E_0(x)$ which can be read off from the
first term of the
divergent part of the energy (\ref{Ediv}):
\begin{equation}
E_0(x)=\frac{\tilde \mu^{2s}}{4\sqrt{\pi}
\Gamma(s-\frac 1 2)} M^{2-2s}
\Gamma (s-1) \, \ \ .\label{Econst}
\end{equation}
Let us now expand $E_0$ around $s=0$:
\begin{equation}
E_0(x)\sim \left( \frac 1s +\gamma_E +1 -\ln
\left( \frac {M^2}{\tilde \mu^2} \right)
-\partial_s \ln \Gamma \left( s-\frac 12 \right)_{s=0}
\right)
\frac {M^2}{4\pi} \, \ \ .\label{Ecs}
\end{equation}
In a renormalizable theory (\ref{Ecs}) is accompanied
by a counterterm
\begin{equation}
\Delta E_0(x)
= -\left( \frac 1s +\gamma_E +1 -\ln
\left( \frac {\nu^2}{\tilde \mu^2} \right)
-\partial_s \ln \Gamma \left( s-\frac 12 \right)_{s=0}
\right)
\frac {M^2}{4\pi} \, \ \ ,
\label{Ect}
\end{equation}
which repeats the pole structure of (\ref{Ecs}).
Finite renormalization
is encoded in an arbitrary parameter $\nu^2$ in
(\ref{Ect}).

Vanishing of tadpoles means that the one-loop
renormalized
energy $E_0(x)+\Delta E_0(x)$ has a minimum at the
same value
of $\Phi =\Phi_0$ as the classical energy.
Generically,
$\partial (M^2)/\partial\Phi \ne 0$ near $\Phi
=\Phi_0$.
We obtain:
\begin{equation}
\frac {\partial}{\partial M^2}(E_0(x)+\Delta E_0(x))
=-
\frac 1{4\pi} \left( \ln
\frac{M^2}{\nu^2} +1\right)=0 \label{notadpole} \ \ .
\end{equation}
This condition defines $\nu^2$ in terms of $M^2(\Phi_0
)$.

Consider now a non-trivial vacuum $\Phi$. The operator
acting
on the fluctuations is modified: $-\partial_1^2
+M(\Phi_0 )^2+V(x)$.
It is essential that $V(x)\to 0$ for $x\to \pm\infty$.
Consider
$E_0^{\rm div}$ (\ref{Ediv}). The first term
containing $a_0$ is now
proportional to $M(\Phi_0 )^2$, which is a constant.
This term
must be removed by renormalization of the
``cosmological constant''.
This is always done in such a way that the whole
infinite volume
contribution coming from $a_0$ is subtracted. The
second term
with $a_{\frac 1 2}$ is simply absent on manifolds without
boundary.
We are left with the 3rd term. It requires a counterterm
\begin{equation}
\Delta E_0=-\left( \frac 1s +\gamma_E +1 -\ln
\left( \frac {\nu^2}{\tilde \mu^2} \right)
-\partial_s \ln \Gamma \left( s-\frac 12 \right)_{s=0}
\right)
\frac{a_1}{2\sqrt{\pi}} \, \ \ .\label{Ekct}
\end{equation}
The fact the such a counterterm is indeed present is
guaranteed
by renormalizability of the theory. As it is enough
to do just
one (mass) renormalization, the terms added to the
pole $\frac 1s$
in the bracket in (\ref{Ekct}) are exactly the same as
in
(\ref{Ect}). Finally we have
\begin{equation}
E_0^{\rm div}+\Delta E_0=-\frac{a_1}{2\sqrt{\pi}}
\left( \ln
\frac{M^2}{\nu^2} +1\right)\, \ \ .\label{E+ct}
\end{equation}
The right hand side of (\ref{E+ct}) vanishes due to
the condition
(\ref{notadpole}). We conclude that the
renormalization condition
that tadpoles vanish in the trivial vacuum is
equivalent to dropping
$E_0^{\rm div}$ in the kink vacuum.

\end{document}